\documentclass{elsart3p}

\usepackage{graphicx}

\usepackage{amssymb}

\begin{document}

\begin{frontmatter}



\title{The bulge in the basal plane area of cuprate superconductors -- 
evidence for $3a$ singlet hole pairs}


\author{J\"urgen R\"ohler}
\address{Fachgruppe Physik, Universit\"at zu K\"oln, 50937 K\"oln, Germany}
\ead{Juergen.Roehler@uni-koeln.de}
\ead[url]{http://www.uni-koeln.de/$\sim$abb12}

\begin{abstract}
The bulge in the doping dependence of the basal plane area in hole doped cuprate 
superconductors is connected with a non-double-occupancy constraint 
for the oxygen cages in the CuO$_2$ lattice. This constraint favors the 
formation of $3a$ hole pairs growing to filaments with gapless excitations along the $(\pi, 
\pi)$ direction. Thus in the pseudogap regime a nodal metal of hole pairs is created. 
Densest packed $3a$ hole pairs stabilize the optimum doped state at $n =1/6 \simeq 0.16$. 

\end{abstract}

\end{frontmatter}


The difficult problem of a theoretical description of the HTSC phase
diagram during the last 20 years has been tackeled by various
scenarios for the doping dependence of the phase lines. These
attacks have been focussed on the electronic properties,
such as the competition between antiferromagnetism and superconductivity, 
but rarely on others, including the special stability of the 
optimum doped state at $n_{opt} = 0.16\pm 0.01$, 
the doping dependences of the interatomic distances in the
electronically operative CuO$_2$ layers. This short note addresses the 
implications for the HTSC phase diagram resulting from the 
recent observation \cite{Roe04a} of a bulge in the doping 
dependence of the basal plane area ($a^2$, or $ab$).

Hole doping of cuprate superconductors removes electrons from the
antibonding $\sigma^* $Cu$3d_{x^2-y^2}$O$2p_{x.y}$ band and thus tends
to decrease the basal plane area on increasing metallization of the
CuO$_ 2$ layers. The basal plane
area is found to shrink by $\simeq 1$ \% between the onset of superconductivity close to the
insulator-metal transition at $n_{I-M}=0.06(2)$ and its disappearence 
in the heavily overdoped regime at $n_{M} = 0.28(2)$. 
But between $n_{sc}=0.06(2)$ and the closure of the large pseudogap at
$n_{PG}=0.21(2)$ its doping dependence is always found concave away from the $n$ axis
instead of convex toward it. This ``bulge'' exhibits a maximum around 
$n_{opt}= 0.16(1)$  and collapses at  $n_{PG}=0.21(2)$, 
notably $within$ the overdoped regime. At optimum doping, 
$n_{opt} = 0.16\pm 0.01$, the basal plane area bulges by up to 30\% relative to the
Pauling-type $1-log$ $m$ contraction expected from the increasing degree of
covalency $m$ on hole doping. Although the CuO$_2$ layers are bound to a 
complex stereochemistry from spacing, insulating and separating 
layers embedding them, their  intrinsic electronic demands are the dominant 
cause in setting the basal plane area. 

Obviously doping up to $n\leq 0.21$ exerts an $outward$ electronic pressure 
on the basal Cu grid counteracting the Pauling-type compressive covalent strain. 
We connect this outward electronic pressure with the organizational 
principles governing the $Aufbau$ of the many body state in 
a hole doped square-planar CuO$_2$ lattice \cite{Roe04b}. 
Holes doped into this particular lattice enter square oxygen cages around each Cu site, 
not simply single Cu--O bonds. Due to the translational invariance of 
the oxygen cage they gain an extra stabilization energy from the resonance between 
the symmetric combination of the four oxygen hole states 
of the cage and the central  Cu$3d^9$ hole. This hole state is known as 
``Zhang-Rice'' (ZR) singlet being generically repulsive. Occupation of 
the oxygen cage with two holes is energetically highly unfavourable. 
Thus the ZR singlet must exclude the four $nn$ oxygen cages 
from occupation with holes and be a ``self-protecting'' singlet (SPS).
We phrase this as the ``second'' non-double-occupancy 
constraint of the HTS problem operative in addition to the large $d-d$ repulsion.

As a consequence the physics of hole doping in this particular type of
a Mott insulator may not be reduced to scenarios of structureless
vacancies moving in an antiferromagnetic background.  Holes in the
CuO$_2$ lattice carry also the symmetry of the resonating singlet state
stabilizing them.  Clearly, such holes tend to be localized.  But
feedback may delocalize them because the repulsion that constrains
their distribution to certain lattice sites is itself determined by
the distribution that it constrains.  Neighbored SPS may connect each
other only in ``site centered'' or ``bond centered'' configurations,
the latter being unstable against the formation of $3a$ hole pairs
\cite{Roe04b}.  Bond centered SPS are connected along the Cu--O bonds
with a Cu--Cu distance of $3a$, thus creating an inversion center for all space and spin coordinates of the
pseudomolecule \cite{Roe04b}.  Notably the $3a$ pairing configurations
are created by a repulsive interaction that relaxes with increasing 
pairing density..

The population of the lattice with $3a$ hole pairs is geometrically
constrained.  Their densest packing occurs at $n = 1/6 \simeq
n_{opt}=0.16$(1).  Most interestingly, the densest packing creates a
perfect electronic mesostructure for a nodal metal of $3a$ hole pairs. 
As visible in Figs.  1, 2 $gapless$ excitations of the $3a$ hole pairs
occur along the zig-zagging filaments along $(\pi,\pi)$.  Fig. 1
displays a snapshot of the mesostructure in the underdoped regime. 
Herein the nodal gapless filaments coexist with a fraction of
localized $3a$ pairs. Overdoping will start to destroy $3a$ pairing,  
but high pairing densities might allow also the formation of bond centered $1a$ pairs.

Summarizing, the outward electronic pressure bulging the basal plane
area in the pseudogap regime is possibly connected with the growth of a
filamentary nodal metal of $3a$ hole pairs.

\begin{figure}[h]

\includegraphics*[width=7.4cm]{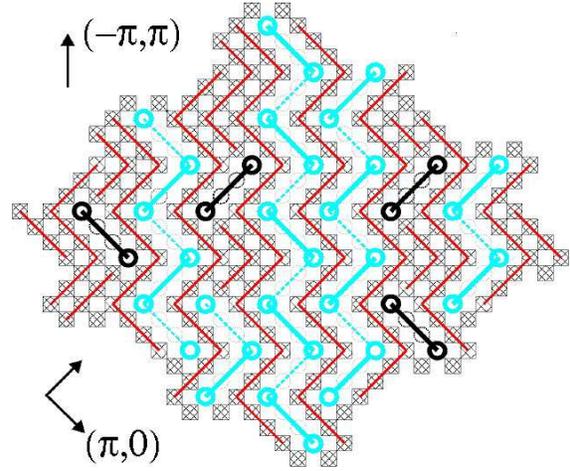}

\caption{Adiabatic snapshot of a underdoped CuO$_2$ lattice $(n <
1/6)$ fractionally occupied by connected $3a$ hole pairs (light) zig-zagging along
the diagonal of the basal plane $(-\pi, \pi)$, and localized $3a$ hole pairs (dark). 
Circles  indicate sites doped with self-protecting hole singlets.
Hatched squares: undoped oxygen cages centered at 
the Cu spin sites. Drawn out lines connecting 
the hatched squares indicate antiferromagnetic spin chains.}\label{ud} \end{figure}

\begin{figure}[h]

\includegraphics*[width=7.4cm]{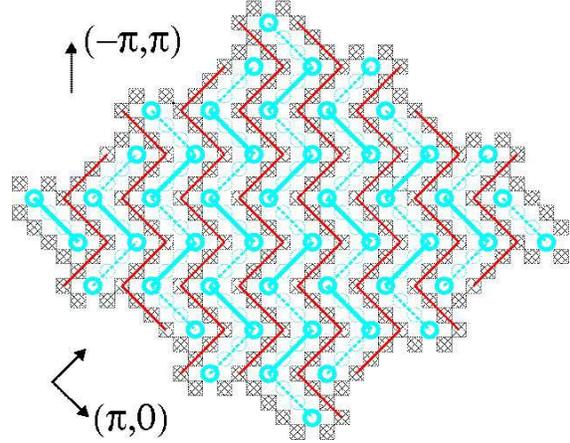}

\caption{Adiabatic snapshot of a optimum doped CuO$_2$ lattice 
most closely packed with $3a$ hole pairs (pairing SPS) at $n_{opt}= 1/6$. 
Connected circles (with weak drawn out/stippled lines):  
$3a$ hole pairs zig-zagging along the diagonal of the basal plane $(-\pi, \pi)$. 
Drawn out and stippled pairs are structurally degenerate. 
Hence connected $3a$ hole pairs are gapless excitations along $(-\pi, \pi)$ 
of nodal metal of hole pairs.} \label{opt}
\end{figure}

\end{document}